# Magnetically recoverable MgFe$_2$O$_4$ nanoparticles as efficient catalysts for rapid dye degradation in water

A. F. Cabrera[1], C. E. Rodríguez Torres[1], P. de la Presa[2,3], S. J. Stewart[1,#]

[1]IFLP-CCT-La Plata-CONICET and Departamento de Física, Facultad de Ciencias Exactas, Universidad Nacional de La Plata, C. C. 67, 1900 La Plata, Argentina

[2] Institute of Applied Magnetism, UCM-ADIF, A6 22,500 km, 28230 Las Rozas, Spain

[3]Department of Materials Physics, Complutense University of Madrid, 28040 Madrid, Spain

[#]Corresponding author: stewart@fisica.unlp.edu.ar

## Abstract

Monophasic MgFe$_2$O$_4$ nanoparticles synthesized by a simple autocombustion method were assessed as magnetically recoverable catalysts for the degradation of methylene blue (MB) in water. The NPs exhibit a crystallite size of ~ 9 nm, a band gap of ~2.11 eV, and soft ferrimagnetic behavior, enabling efficient photocatalytic and Fenton-like activity. The effects of irradiation, H$_2$O$_2$ concentration, agitation mode, catalyst loading, and exposure time were systematically evaluated. Rapid and complete MB discoloration was achieved within minutes in the presence of H$_2$O$_2$, even without illumination, indicating that the process is dominated by a surface-mediated heterogeneous Fenton-like mechanism rather than photocatalysis. Kinetic analysis reveals pseudo-first-order behavior, with rate constants governed by the combined effects of catalyst concentration, oxidant dosage, and dye concentration. Structural stability and excellent recyclability confirm the robustness of the catalyst. These findings position MgFe$_2$O$_4$ nanoparticles as a low-cost, efficient, and reusable material for sustainable wastewater under operationally simple conditions.

# 1. Introduction

The release of synthetic dyes and pigments into natural water systems remains an environmental challenge. Effective remediation requires processes that are environmentally benign, economically viable, and scalable, in accordance with the principles of green chemistry [1, 2]. Among various treatment strategies, advanced oxidation processes (AOPs) have gained prominence for the degradation of dye-derived organic contaminants due to their ability to generate highly reactive oxygen species (ROS)—particularly most notably hydroxyl radicals—which mineralize organic pollutants into $CO_2$, $H_2O$, and inorganic ions [1, 3]. A representative example is heterogeneous photocatalysis, where a semiconductor photocatalyst activated by UV and/or visible light produces reactive radicals capable of decomposing organic dyes. Another widely used AOP for the destruction of organic pollutants in water is the homogeneous Fenton reaction, which involves the catalytic activation of hydrogen peroxide ($H_2O_2$) by dissolved $Fe^{2+}$ ions to generate hydroxyl radicals through a redox cycle ($Fe^{2+}/Fe^{3+}$). This homogeneous process is highly efficient but typically requires acidic conditions (pH ~2.5–3.5) and generates iron-containing sludge as a by-product limiting its practical implementation [4]. In contrast, heterogeneous Fenton or Fenton-like reactions, utilize $Fe^{3+}$ species in solution or, more commonly, solid iron-containing catalysts— such as iron oxides and ferrites—to activate $H_2O_2$ at the catalyst surface. In these systems, the heterogeneous catalyst surface promotes the generation of ROS through $Fe^{3+}/Fe^{2+}$ surface redox cycling, enabling operation under milder pH conditions, improving catalyst recovery, and reducing secondary waste generation.

Spinel ferrites with the general composition $MFe_2O_4$ (M = divalent metal) constitute a versatile family of semiconducting or insulating materials whose optical and magnetic properties are governed by the cation distribution between tetrahedral and octahedral sites. Their low toxicity, chemical robustness, and tunable physicochemical characteristics have motivated their use in catalysis and photocatalysis [5, 6, 7]. The ability of ferrite materials to activate hydrogen peroxide broadens their applicability in both photocatalytic and photo-Fenton degradation pathways. Importantly, ferrimagnetic ferrites can be easily recovered from aqueous suspensions using an external magnetic field, enabling catalyst reuse and reducing operational costs.

Within this family, magnesium ferrite ($MgFe_2O_4$) has attracted considerable attention owing to its moderate band gap (~2.2 eV), high stability in aqueous media, and magnetic response that allows facile solid–liquid separation [8, 9, 10, 11]. Its low production cost, straightforward synthesis, and non-toxic nature further position $MgFe_2O_4$ as an appealing material for sustainable water purification technologies.

Several studies have demonstrated the photocatalytic activity of $MgFe_2O_4$ toward dye degradation, particularly methylene blue (MB) [8, 9, 12, 13, 14]. Depending on the synthesis method the resulting nanostructures exhibit variable efficiencies, often enhanced by the presence of hydrogen peroxide that acts as an oxidant. Shahid et al. [8] showed that $MgFe_2O_4$ synthesized by solid-state reaction exhibits photocatalytic activity toward MB under visible light. Nanoparticles produced via the glycine–nitrate method displayed high degradation efficiency in the presence of $H_2O_2$ [12], while those obtained by solution combustion using urea achieved comparable performance under light irradiation with added $H_2O_2$ [13]. We have previously reported that nanostructured Mg-ferrites prepared by autocombustion or polymerization routes effectively catalyze MB degradation; in these systems, carbonaceous surface residues have been shown to modulate electron availability and suppress electron–hole recombination, thereby improving catalytic performance [9]. More recently, Bus et al. 14] demonstrated that microwave plasma treatment of sol–gel–derived $MgFe_2O_4$ nanoparticles substantially improves their photocatalytic activity toward dyes under sunlight irradiation. Most of these works primarily emphasize photoactivated pathways, while the relative contribution of surface-mediated heterogeneous Fenton-like reactions—particularly under dark or low-energy conditions—remains less systematically explored. As a result, the operational relevance of $MgFe_2O_4$ as a catalyst capable of efficient dye degradation without external energy input is still not fully established.

In the present work, we conduct a comprehensive investigation of the catalytic activity of nanostructured $MgFe_2O_4$ synthesized by the autocombustion method focusing on its performance on the MB degradation in aqueous solution under a wide range of operational conditio. We examine the influence of irradiation conditions, $H_2O_2$ concentration, agitation mode (mechanical stirring, ultrasound, or static), catalyst-to-dye ratio, exposure time, and catalyst reusability. The results provide new insight into the operational parameters governing $MgFe_2O_4$-based AOPs and further establish its potential for cost-effective, energy-efficient and magnetically recoverable catalysts for sustainable wastewater treatment.

## 2. Experimental

### 2. 1 *Synthesis and characterization of nanoferrites $MgFe_2O_4$ catalysts*

Nanosized $MgFe_2O_4$ ferrites (NPs) were synthesized by the autocombustion method [9-CAB20]. Structural, microestructural, magnetic and optical characterization was carried out by X-ray diffraction (Cu Kα, λ = 1.5406 Å), high resolution transmission electron microscopy (HRTEM), $^{57}Fe$ Mössbauer spectroscopy, DC magnetization measurements, and diffuse reflectance spectroscopy in the UV–Vis range (see Supporting Information (S. I.) ). Additionally, Fourier Transform Infrared (FTIR) spectroscopy and X-ray Photoelectron Spectroscopy (XPS) were employed to identify surface functional groups and to assess the chemical stability of the synthesized samples before and after their use in the degradation experiments described below. FTIR measurements were performed using a Nicolet iS20 FTIR spectrometer (Thermo Scientific) operating in the spectral range of 4000–400 $cm^{-1}$ with a resolution of 4 $cm^{-1}$. Each spectrum represents the average of 16 accumulated scans recorded in transmission mode at room temperature. XPS analyses were conducted using an XR50 X-ray source (SPECS) equipped with Al Kα (1486.6 eV) and Mg Kα (1253.6 eV) anodes, operated at 13 kV and 300 W, and a PHOIBOS 100 MCD hemispherical analyzer (SPECS). Measurements were performed under ultra-high vacuum conditions (0.9–3 × $10^{-8}$ mbar). Prior to data acquisition, the energy scale was calibrated using gold and copper foils cleaned with an $Ar^+$ ion gun, taking Au $4f_{7/2}$ at 84.00 eV and Cu $2p_{3/2}$ at 932.66 eV as reference peaks. Elemental quantification, expressed in atomic percent, was performed using relative sensitivity factors and assuming a homogeneous elemental distribution within the analyzed area (2 × 5 mm²). Hydrodynamic particle size distribution measurements were carried out using a Horiba Partica LA-950 V2 laser diffraction (LD) analyzer. The instrument operates using both Mie theory and the Fraunhofer optical model, enabling accurate determination of particle size distributions for samples in suspension. Measurements were performed with the NPsdispersed in water, ensuring proper dilution and preventing multiple scattering effects. All analyses were conducted at room temperature on NPs before and after their use in the degradation of methylene blue following the standard operating procedures recommended by the manufacturer.

*2. 2 Catalytic degradation tests*

The degradation of methylene blue (MB) dye with the addition of NPs under different conditions (see details below) was tested using UV-vis absorption. After a certain interval of time an aliquot of the liquid was collected via centrifugation and placed in a spectrophotometer cell, after decanting the suspended magnetic particles using a magnet. The degradation efficiency percentage (DE) was estimated from DE% = 100 x (1 - $I_{MBts}/I_{MB}$), where $I_{MBts}$ and $I_{MB}$ are the highest peak intensities of the spectra of the supernatant of MB plus catalyst mixtures and pure MB, respectively, after a certain time interval ($t$). The degradation studies were conducted with and without visible light illumination, using a 9 W light-emitting diode lamp positioned 20 cm above the liquid surface.

To evaluate the catalytic activity of $MgFe_2O_4$ in the degradation of MB, the following tests were conducted:

*Test A*: *Influence of hydrogen peroxide and illumination*: Aqueous suspensions (total volume of 15 mL) containing MB (12 mg.L$^{-1}$; pH 7.05) were mixed with the NPs (0.8 g.L$^{-1}$) for 30 minutes using a sonicator in the presence and in the absence of hydrogen peroxide ($H_2O_2$) (75 g.L$^{-1}$, added as 5 mL of a 30 wt% $H_2O_2$), with and without illumination. The incidence of the sequential addition of NPs and $H_2O_2$ was also checked.

*Test B*: *Effect of mixing conditions*: To evaluate the importance of NPs dispersion in the catalytic process, aqueous solutions containing MB were mixed with the NPs and $H_2O_2$ (all components in the same concentrations used in test A) during 30 minutes using either ultrasound (US), a mechanical stirring (MS) or leaving the mixture in static conditions (SC) for 30 min., with and without illumination.

*Test C*: *Effect of peroxide concentration on MB degradation in dark conditions:* To further study the influence of $H_2O_2$, aqueous solutions of MB (12 mg.L$^{-1}$) were mixed with the NPs (0.8 g.L$^{-1}$) for 30 minutes using US with different concentrations of $H_2O_2$, without illumination.

*Test D*: *Kinetics study:* Aqueous solutions containing MB, NPs and $H_2O_2$ in the same amounts used in test A, were mixed using US, without illumination. Every five minutes an aliquot of the liquid was separated and placed in a spectrophotometer cell, to complete a total interval of time of 40 minutes.

*Test E: Reusability of NPs*: The NPs were reused under the same conditions described for Test A, namely ultrasound-assisted mixing, with and without illumination. The reused NPs were employed directly, without any intermediate washing or regeneration step.

*Test F: Natural light-assisted MB degradation:* Aqueous MB suspensions (1.3 and 12 mg L$^{-1}$, with initial pH values of 7.28 and 7.05, respectively) were brought into contact under static conditions with $MgFe_2O_4$ nanoparticles (0.8 and 1.6 g L$^{-1}$) and varying volumes of $H_2O_2$ for exposure periods of up to 8 or 24 h. The samples were maintained under natural indoor illumination, experiencing alternating day–night cycles, in order to evaluate catalyst performance under low-energy and environmentally relevant conditions. This experimental design allows the individual and combined effects of light irradiation, oxidant activation, mixing conditions, and catalyst reuse to be decoupled, providing a comprehensive assessment of the catalytic performance and operational robustness of $MgFe_2O_4$ nanoparticles.

Table I: Summary of the experimental conditions used to evaluate the degradation of methylene blue (MB) dye. NPs: as-prepared Mg-ferrite nanoparticles, US: ultrasound, MS: mechanical stirring, SC: static conditions. Each experiment is described in the text. The pH corresponds to the initial solutions with catalysts.

| Test | MB (mg.L$^{-1}$) | NPs (g.L$^{-1}$) | H$_2$O$_2$ (mL) | Mixing | Illumination | Time | pH | Others |
|---|---|---|---|---|---|---|---|---|
| A | 12 | 0.8 | 0 and 5 | US | Yes/No | 30 min | 9.25; 5.38 | |
| B | 12 | 0.8 | 5 | US; MS; SC | Yes/No | 30 min | 5.38 | sequence NPs/H$_2$O$_2$ |
| C | 12 | 0.8 | 0.6; 1; 2.5; 5 | | No | 30 min | 7.4; 7.17; 6.2; 5.38 | |
| D | 12 | 0.8 | 5 | US | | 5-40 min | 5.38 | |
| E | 12 | 0.8 | 0 and 5 | US | Yes/No | 30 min | 9.25; 5.38 | reuse |
| F | 1.3/12 | 0.8/1.6 | 0 to 5 | SC | Sunlight | 8/24 h | 8.4/5.07 | |

## 3. Results and Discussion

### 3.1 *Characterization of the MgFe$_2$O$_4$ catalyst prepared by autocombustion*

XRD results indicate that the MgFe$_2$O$_4$ nanoparticles (NPs) synthesized by the autocombustion method crystallize in the spinel structure, with an average crystallite size of approximately 9 nm and a lattice parameter of *a* = 8.39 Å. The average particle diameter determined by TEM (D = 10 ± 2 nm; see SI) is in good agreement with the XRD results, and HRTEM images confirm the high crystallinity of the ferrite nanoparticles. Similar TEM and HRTEM features were observed after their use as catalysts in the discoloration of MB (see SI), indicating preserved structural integrity. LD measurements reveal that the NPs form micrometric aggregates with a broad size distribution (see Section 3.5). UV–vis spectroscopy yields an optical band gap of approximately 2.11 eV. Mössbauer spectroscopy indicates that iron is predominantly in the Fe$^{3+}$ oxidation state, consistent with the formation of Mg-ferrite nanoparticles exhibiting mainly superparamagnetic relaxation at room temperature. The magnetic response corresponds to that of a soft ferrimagnetic material, with a magnetization of 11 A·m²·kg⁻¹ under an applied field of 2 T. Additional characterization details are provided in the Supplementary Information, while FTIR, XPS, and LD results are discussed in Sections 3.3–3.5.

3. 2 *Catalytic results*

*3.2.1 Results of test A*: *Enhanced catalytic performance of $MgFe_2O_4$ catalyst in MB removal: role of $H_2O_2$, illumination and addition sequence.*

The DE of methylene blue degradation was evaluated by comparing four systems (Fig. 1(a)) labelled: (i) MB + $H_2O_2$, (ii) MB + NPs, and (iii) MB + NPs + $H_2O_2$. As shown in Figure 1(b), the absorbance spectrum of MB in the optical range exhibits two characteristic peaks at approximately 609 and 665 nm [9, 15]. When only $H_2O_2$ is added, the degradation reaches about 20%, regardless of illumination, after 30 min of ultrasonic treatment, confirming its limited oxidation ability without catalyst activation. In contrast, when only Mg-ferrite nanoparticles are used as the sole active component, a pronounced decrease in absorption is observed, corresponding to a DE of nearly 58% (Fig. 1 (a)). This result does not vary significantly under illumination. The addition of $H_2O_2$ to the MB + NPs system produces a remarkable enhancement in catalytic performance, achieving complete discoloration (DE values of 100% and 93% with and without illumination, respectively) within 30 minutes. The negligible influence of light indicates that the process is primarily governed by a surface mediated heterogenous Fenton-like mechanism rather than a photocatalytic one. The efficient activation of $H_2O_2$ is attributed to the $Fe^{3+}/Fe^{2+}$ redox couple at the ferrite surface, as evidenced by XPS analysis, results of which are presented below (section 3.4). These findings highlight, on one hand, the crucial catalytic role of the NPs in MB removal and, on the other, the synergistic effect of both components NPs and $H_2O_2$ that enable complete degradation within a relatively short time. It is also noteworthy that, in the latter case, the outcome was independent of the order in which the NPs and $H_2O_2$ were added to the solution (see Fig. 1 (b)).

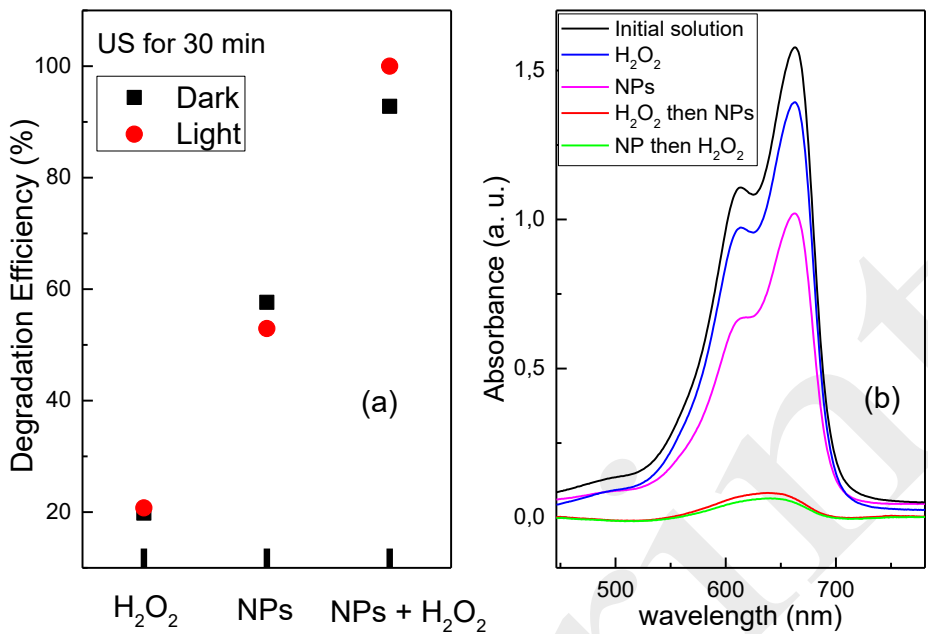

**Figure 1:** (a) Degradation Efficiency (DE) percentage of MB aqueous solution with (Light) and without (Dark) visible light irradiation for test A described in the text. (b) Representative absorbance spectra recorded in the absence of visible light irradiation.

*3.2.2 Results of test B*: *Exploring the role of NPs degree of dispersion*

Considering that the best performance in test A was achieved by combining NPs and peroxide, regardless of the order of their addition to the solution, the discoloration reaction using these components was studied under three different mixing conditions: ultrasonic agitation (US), mechanical stirring (MS) and static conditions (SC), over the same time interval. As shown in Figure 2, the DE values increase following the sequence SC → MS → US, indicating that higher dispersion of the NPs enhances the catalytic efficiency. This result indicates that mass transport, particle dispersion and intimate contact between reactants and active sites are crucial for efficient oxidation. The enhanced catalytic activity under US is attributed to improved accessibility of MB molecules to reactive surface sites and to the continuous renewal of the catalyst–solution interface. Moreover, no significant changes are observed under illumination, supporting the idea that the process is mainly governed by catalytic Fenton-like reactions rather than by photoactivation. The results are consistent with the surface-controlled mechanism inferred from FTIR and XPS analyses for samples before and after the reactions, which are discussed in more detail in later sections of this work.

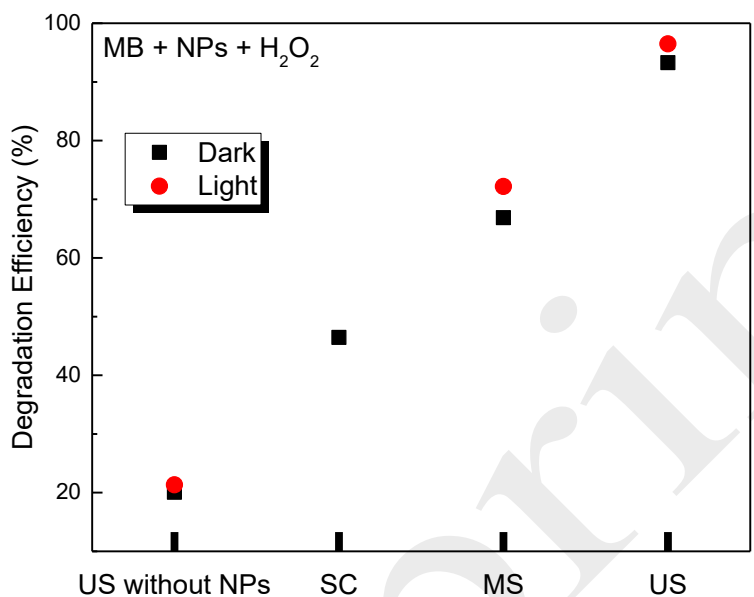

**Figure 2:** Degradation efficiency (DE) percentage with and without visible light irradiation of MB+NPs+ $H_2O_2$ system (test B) during 30 min. The horizontal axis indicates the method of mixture, which are SC: static conditions; MS mechanical stirring; US ultrasound as well as the DE of a solution only adding $H_2O_2$ (US without NPs).

*3.2.3 Results of test C: Optimization of $H_2O_2$ concentration for enhanced MB degradation with NPs*

The catalytic activity of the $MgFe_2O_4$ nanoparticles was further examined as a function of hydrogen peroxide concentration under identical conditions of reaction time, absence of illumination, and constant nanoparticle loading. As shown in Figure 3, increasing the volume of $H_2O_2$ initially leads to a marked enhancement in DE, consistent with the greater availability of oxidant to generate ROS.

However, once the H₂O₂ volume exceeds 1 mL, the incremental improvement in DE becomes significantly diminished. This behavior aligns with the well-established scavenging effect of excess H₂O₂, whereby surplus oxidant reacts with •OH radicals—rather than with the target dye—thereby reducing the overall degradation efficiency. This behavior reflects the dual role of H₂O₂ in Fenton-like systems: at optimal levels, it acts as an efficient source of •OH radicals through the redox reactions [16]

$$Fe^{3+}(surface) + H_2O_2 \rightarrow Fe^{2+}(surface) + \bullet OOH + H^+$$

$$Fe^{2+}(surface) + H_2O_2 \rightarrow Fe^{3+}(surface) + \bullet OH + OH^-$$

Beyond this optimum, excessive oxidant doses lead to radical consumption and a net decrease in catalytic performance. The observed trend supports a heterogeneous Fenton-type mechanism in which the balance between surface $Fe^{2+}/Fe^{3+}$ species—also corroborated by XPS analyses—governs the degradation efficiency and defines the optimal operating window for effective MB removal.

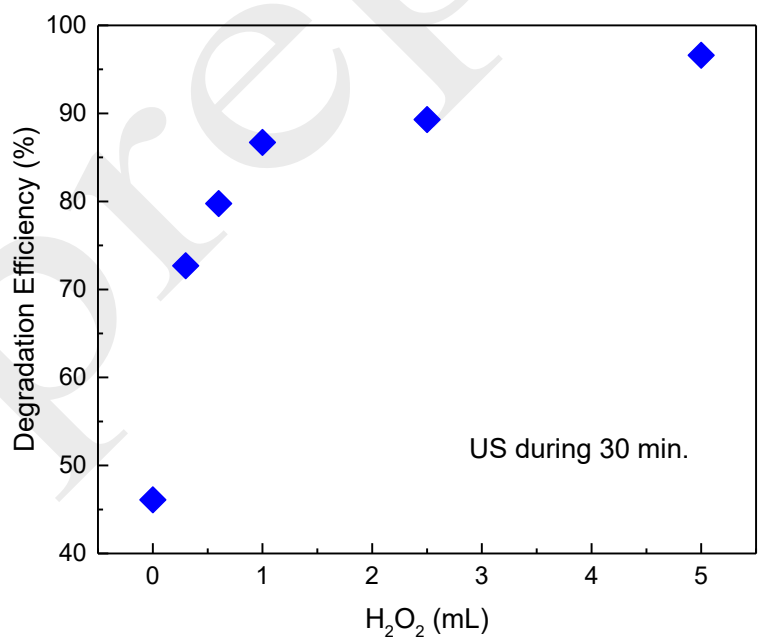

**Figure 3:** Degradation efficiency (DE) percentage using MB+ NPs+ H₂O₂ system mixed using ultrasound for 30 min for different amounts of H₂O₂ (test C).

*3.2.3 Results of test D: First-order degradation kinetics*

Figure 4 shows the degradation of MB as a function of time after the addition of both $H_2O_2$ and NPs. After the first five minutes of ultrasound, the DE reaches approximately 60%, and exceeds 80% after 10 minutes. The rate of MB degradation was checked with a pseudo-first-order model

$$\ln\left(\frac{C_t}{C_0}\right) = -kt \qquad \text{eq. (1)}$$

where $C_0$ is the initial concentration of MB, $C_t$ is the concentration after $t$ time and $k$ the rate constant, being the latter $k = 0.0110(4)$ min$^{-1}$ (see inset Fig. 4). This shows the good catalytic efficiency of the Mg-ferrite NPs in activating $H_2O_2$, leading to a rapid generation of ROS that accelerates the degradation process. The enhanced surface oxygen content (see subsection 3.3) likely facilitates electron transfer and peroxide activation, further supporting the synergistic interaction between the NPs and $H_2O_2$ in achieving fast and efficient MB degradation.

Thus, tests A to D show that $MgFe_2O_4$ works predominantly as a heterogeneous Fenton-like catalyst, enabling rapid $H_2O_2$ activation without external energy input. The reaction is controlled by surface redox processes and is only weakly affected by illumination, while nanoparticle dispersion and oxidant concentration dictate the overall efficiency.

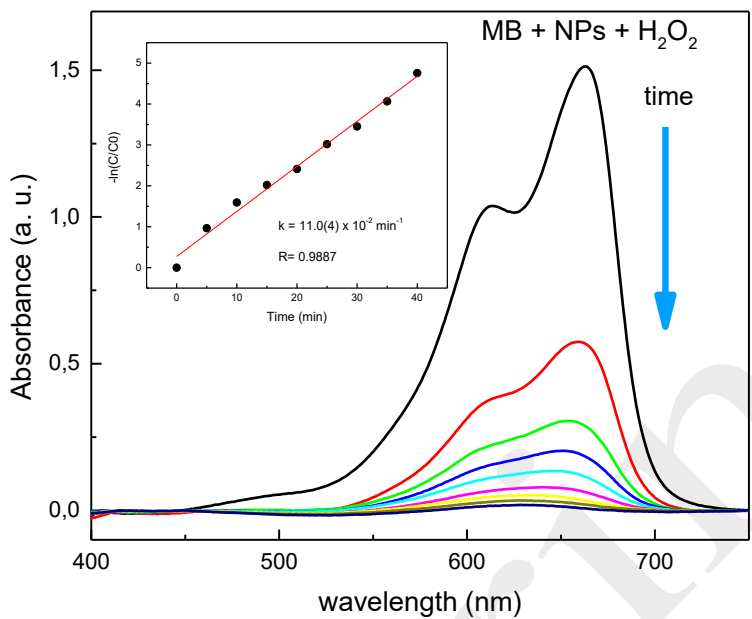

**Figure 4**: Visible absorbance spectra corresponding to test E performed on MB aqueous solution with addition of NPs and $H_2O_2$. The arrow points towards the progression of time as the spectra were recorded. Inset: Linear regression is used to determine the rate constant *k* in kinetic reaction.

*3.2.4 Results of test E: Reusability of NPs in MB Degradation*

Reusability tests were performed to evaluate the structural stability and operational durability of the $MgFe_2O_4$ catalyst. After each 30-min degradation cycle, the nanoparticles were magnetically recovered and reused under identical experimental conditions, without washing, drying, or any additional pretreatment between cycles. This aspect is particularly relevant since in many previously reported studies the catalyst is subjected to washing and thermal drying steps prior to reuse, which may partially restore surface activity or remove adsorbed reaction intermediates. For instance, Wang *et al.* [17] reported that Prussian-blue-modified magnetic nanoparticles were washed three times with ultrapure water and dried at 60 °C before each reuse cycle, while Luong *et al.* [18] recovered their CuCo-ZIF catalyst by centrifugation, followed by multiple ethanol washing steps and drying at 60 °C for 24 h.

Similarly, in the work of Silva *et al.* [19] recovered their $CuFe_2O_4$–$Fe_2O_3$ catalyst by magnetic separation, washed it with distilled water and dried it at 100 °C for 24 h between reuse cycles.

In contrast, the present protocol deliberately avoids any regeneration procedure, providing a more stringent evaluation of catalyst durability under realistic operating conditions. When only NPs are employed, their catalytic efficiency decreases markedly upon reuse: the DE reaches 56% in the first cycle but drops to 20% and 17% in the second and third cycles, respectively (Fig. 5). This rapid deactivation is attributed to the accumulation of adsorbed dye molecules and partially oxidized intermediates on the catalyst surface, which progressively block the active Fe sites. Illumination does not significantly modify this trend. In contrast, when $MgFe_2O_4$ nanoparticles are used together with $H_2O_2$, the DE remains high during the first two cycles under illumination, indicating that the synergistic interaction between the ferrite surface and the oxidant not only enhances the initial degradation rate but also improves short-term catalytic stability. Nevertheless, a decrease in performance is observed in the third cycle (DE ≈ 56%), suggesting partial surface deactivation or loss of active sites. Under dark conditions, the deterioration occurs more rapidly, and the DE falls to ~20% by the third reuse. These results demonstrate that $MgFe_2O_4$ nanoparticles retain good short-term reusability without any regeneration treatment, particularly in the presence of $H_2O_2$, which effectively mitigates surface deactivation during repeated cycles.

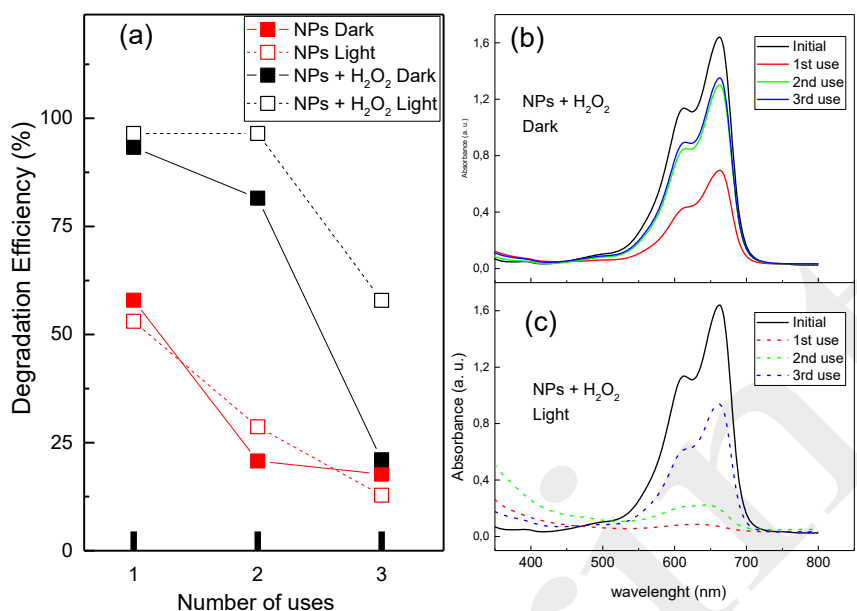

**Figure 5:** (a) Degradation efficiency percentage of MB aqueous solution with the NPs catalysts, with and without $H_2O_2$ during 30 min of ultrasound, with (Light) and without (Dark) illumination (test E). (b) and (c) Absorbance spectra corresponding to MB aqueous solution with NPs + $H_2O_2$ under darkness and illuminated conditions.

*3.2.5 Results of test F: Natural-light-driven degradation of methylene blue under mild conditions.*

This experiment highlights the low-cost and energy-efficient character of the degradation process. The reaction proceeds exclusively under natural light during a day-night cycle, without the need for artificial illumination and without any external mixing, operational costs and energy requirements are minimized. Figure 6 shows the visual evolution of the MB degradation. The decoloration observed in the presence of NPs demonstrates their catalytic activity under mild conditions, while the decantation of the particles reflects their aggregation and facilitates post-treatment recovery. This effect was also observed even at low initial concentrations (i. e., using MB= 1.3 mg L$^{-1}$, not shown).

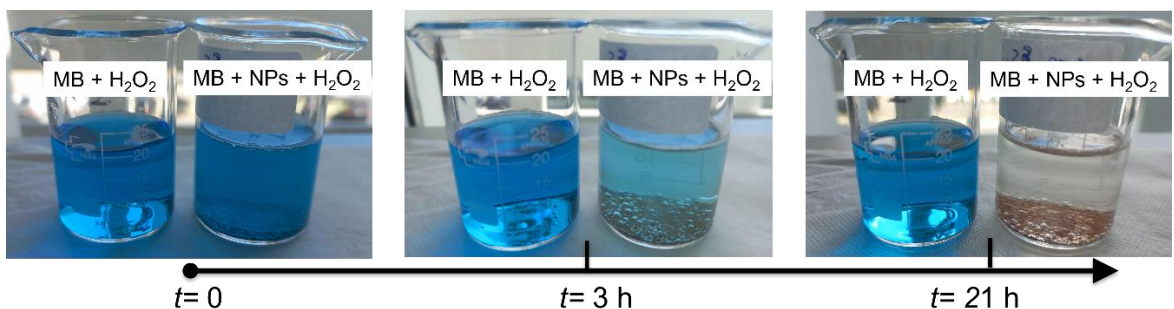

**Figure 6:** Results of the methylene blue (MB) degradation under natural light (test F). Photographs show the MB + $H_2O_2$ system in the absence and in the presence of the $MgFe_2O_4$ catalyst (NPs) at the initial time and after representative reaction times of 3 and 21 h, as indicated by the time arrow. This illustrates the results using 5 mL of $H_2O_2$ and 0.8 g.L$^{-1}$ of NPs.

These experiments carried out using different amounts of $H_2O_2$ show that the rate of MB degradation reaction follows a pseudo-first order kinetics, with $k$ values of 0.0043(2), 0.0042(2), 0.0062(2) min$^{-1}$ for 0.6, 1 and 5 mL of $H_2O_2$, respectively. These results would indicate that increasing the volume of $H_2O_2$ limits the generation of radicals. On the other hand, duplicating the amount of NPs accelerates the reaction, being $k = 0.008(5)$ min$^{-1}$, confirming that the reaction rate is governed by the availability of active catalytic surface sites. The reaction proceeds efficiently despite the absence of forced mixing, indicating that surface-mediated Fenton-like reactions remain operative under environmentally relevant conditions. These results confirm the catalyst's stability and reinforce its potential for environmentally oriented applications that rely on natural irradiation and mild operating conditions.

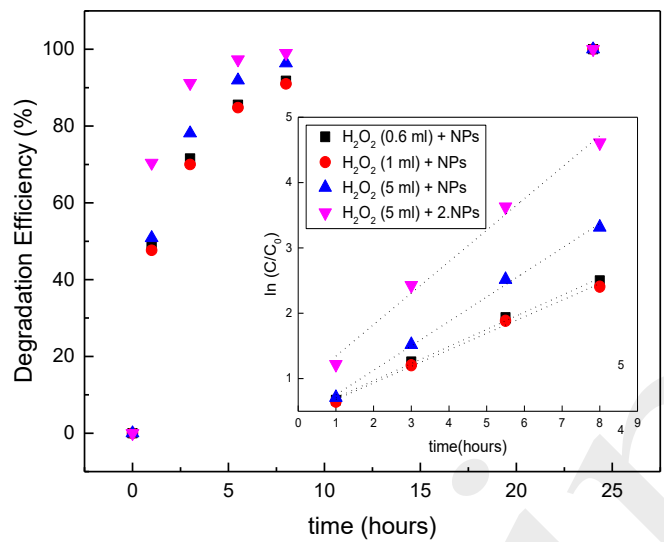

**Figure 7:** Evolution of the MB degradation under natural light without external mixing using NPs and $H_2O_2$ in different proportions as described in test F. Inset: Linear regression to fit the empirical data to determine the rate constants $k$ of the kinetic reaction.

The degradation experiments carried out under natural sunlight and static conditions (Test F) are fully consistent with the general kinetic framework summarized in Figure 8. As shown in Figure 8(a), $k$ increases proportionally with product between the nanoparticle concentration [NP] and the number of moles of hydrogen peroxide ($n(H_2O_2)$), consistent with a heterogeneous Fenton-like mechanism in which the kinetics are governed by the density of catalytic surface sites and the oxidant flux responsible for •OH generation. It is worth noting that within the dataset corresponding to 12 mg $L^{-1}$ of MB, one data point was obtained using twice the nanoparticle concentration (1.6 g.$L^{-1}$), whereas the remaining measurements were performed with 0.8 g.$L^{-1}$ and varying peroxide concentrations. This additional point was included to better illustrate the dependence of the rate constant on the total amount of catalytic material.

When $k$ is plotted as [NP] × $n(H_2O_2)$ / $n(MB)$ (Fig. 8(b)), where $n(MB)$ is the number of mol of MB, the data obtained at 1.3 and 12 mg $L^{-1}$ MB fall on separate linear trends, indicating that the reaction rate depends explicitly on the initial dye concentration. However, normalization by $[n(MB)]^{3/2}$ (Fig. 8(c)) collapses all data onto a single linear relationship, revealing an effective reaction order in MB close to 3/2. Such a fractional-order dependence is consistent with conditions in which only the adsorbed MB fraction is reactive, while adsorption equilibria and/or the coexistence of monomeric and aggregated MB species introduce a non-linear response to the bulk MB concentration. The master-curve behavior obtained with $[n(MB)]^{3/2}$ suggests that the degradation rate reflects coupled processes of adsorption, surface reaction, and radical propagation, which collectively produce the observed mixed-order kinetics. This effective reaction order should be regarded as a phenomenological descriptor, reflecting the overall kinetic behavior, rather than a mechanistic or stoichiometric reaction order.

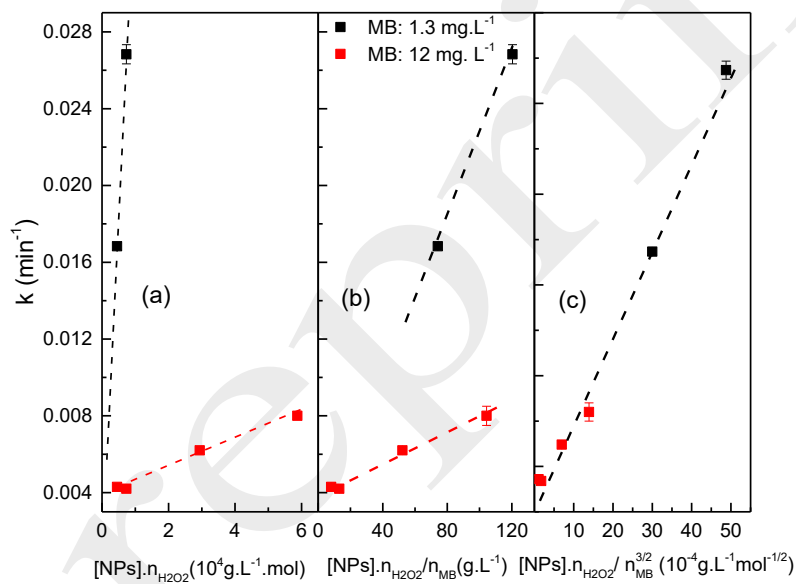

**Figure 8:** Pseudo–first-order rate constants ($k$) for MB degradation as a function of (a) [NP] × $n(H_2O_2)$, (b) [NP] × $n(H_2O_2)$ / $n(MB)$, and (c) [NP] × $n(H_2O_2)$ / $[n(MB)]^{3/2}$. The collapse of all data into a single trend in (c) indicates an effective MB reaction order of ~3/2.

3.3 *FTIR analysis of MgFe$_2$O$_4$ catalyst before and after the degradation reaction.*

The FTIR spectrum of MgFe$_2$O$_4$ nanoparticles exhibits the two intense absorption bands located around 580 cm$^{-1}$ and 420 cm$^{-1}$ (Fig. 9 (a)), attributed to the stretching vibrations of the metal–oxygen bonds at tetrahedral and octahedral sites, respectively—features characteristic of the spinel ferrite structure [20]. Similar spectra were recorded after subjecting the NPs to a thermal treatment at 600 °C (labelled NPs + T600) or ultrasonic dispersion in water with and without peroxide (NPs + w and NPs + w + $H_2O_2$, respectively) (Fig. 9 (a)), indicating that the spinel structure remains stable under these conditions. On the other hand, broad absorption features centred between 2700 and 3600 cm$^{-1}$ correspond to O–H stretching vibrations of surface hydroxyl groups, while the peaks near 1600 and 1390 cm$^{-1}$ (Fig. 9 (a)) are assigned to the stretching modes of carboxyl and nitrate groups, respectively [21, 22, 23]. These residual groups, resulting from incomplete decomposition of organic precursors during the combustion process [9], are gradually reduced by thermal (NPs+T600) or ultrasonic treatments (NPs+w) (see Fig. 9(a)). A band at ~860 cm$^{-1}$ corresponds to C–H deformation vibrations from trace organic residues, which also diminish upon heating.

Figure 9(b) shows the FTIR spectra of $MgFe_2O_4$ samples before (NPs) and after catalytic degradation of MB during 30 min., with and without $H_2O_2$, labelled MB+NPs and NPs+MB+$H_2O_2$, respectively (samples after test A). In the 3600–2200 cm$^{-1}$ region, variations in the intensity of hydroxyl-related bands indicate surface modification upon reaction. This feature is weaker for the NPs+MB spectrum rather than NPs+MB+$H_2O_2$ one, suggesting that the presence of peroxide enhances surface hydroxylation.

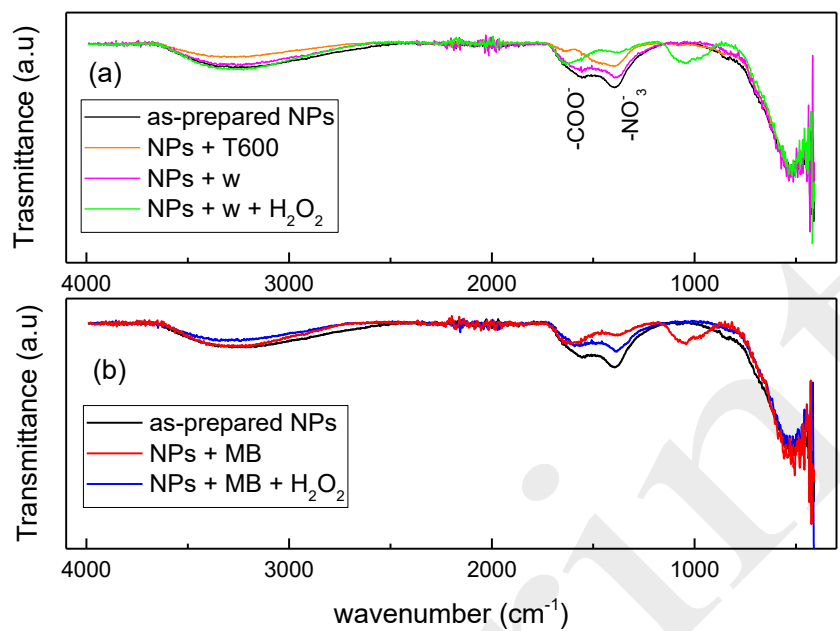

**Figure 9:** (a) ) FTIR spectra of magnesium ferrites: as-prepared (NPs), after thermal treated at 600 °C (NPs + T600), after ultrasonic dispersion in water for 30 min. with and without $H_2O_2$ (NPs + w and NPs + w + $H_2O_2$, respectively); (b) FTIR spectra of magnesium ferrites: as-prepared NPs, after use with MB under ultrasound treatment (NPs + MB), and after MB degradation with addition $H_2O_2$ (NPs + MB + $H_2O_2$) under ultrasound treatment for 30 min.

More pronounced changes occur in the 1760–1170 cm$^{-1}$ range, where the relative intensity of carboxyl and nitrate peaks decreases after the reaction, especially in peroxide-treated samples. This effect suggests partial consumption of these groups during the oxidation process, consistent with their role as adsorption or reaction sites for the dye molecules. Additionally, a new feature emerging around 1100–850 cm$^{-1}$—also detected in the sample NPs + w + $H_2O_2$ (Fig. 9 (a))— can be attributed to intermediates that appears as a consequence of the interaction between $H_2O_2$ and the organic surface residues product of the NPs synthesis [9, 24, 25].

The results demonstrate that MgFe$_2$O$_4$ nanoparticles act as efficient and stable Fenton-like catalysts for the degradation of MB under mild conditions. The stability of the tetrahedral and octahedral Fe–O vibrations confirm the structural robustness of the spinel framework throughout repeated catalytic cycles. The observed evolution of hydroxyl and nitrate surface groups upon $H_2O_2$ treatment indicates active surface modification, which facilitates MB adsorption and promotes subsequent oxidation reactions. In addition, the emergence of a new infrared band near 1100 cm$^{-1}$ suggests the transient formation of intermediate oxidation species, due to $H_2O_2$ and NPs surface residues interaction, providing a direct spectroscopic link between the surface chemistry and the catalytic functionality of the ferrite.

Overall, these findings highlight the chemical and structural resilience of $MgFe_2O_4$, its strong interaction with peroxide, and its potential as a reusable and environmentally sustainable catalyst for wastewater treatment applications.

3.4 *XPS results of MgFe$_2$O$_4$ catalyst before and after the degradation reaction.*

Figure 10 (a) presents the Fe 2p XPS spectra of pristine $MgFe_2O_4$ (NPs) and after being used in the MB discoloration, i. e., samples NPs+MB and NPs+MB+$H_2O_2$. The spectra clearly reveal the signals corresponding to $Fe^{3+}$ in distinct lattice environments. The contribution around 711-714 eV is assigned to $Fe^{3+}$ $2p_{3/2}$ in spinel sites. A satellite feature appears at 719 eV. The $Fe^{3+}$ $2p_{1/2}$ peaks are located at 725.16 eV and 728.4 eV, accompanied by a satellite at 733.5 eV [26].

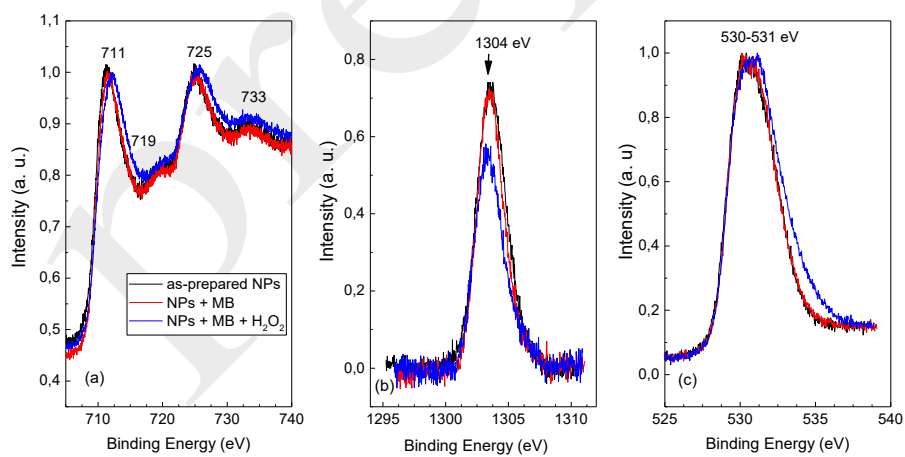

**Figure 10:** XPS spectra of $MgFe_2O_4$ nanoparticles in the as-prepared state and after use in methylene blue (MB) discoloration, both in the absence and in the presence of $H_2O_2$: (a) Fe 2p core-level spectra, (b) Mg 1s core-level spectra, and (c) O 1s core-level spectra for the different samples.

On the other side, the Fe 2$p$ spectrum of NPs+MB+$H_2O_2$ closely resembles that of pristine NPs, although a slight shift toward higher binding energies is observed for the peroxide-treated sample. This shift can be attributed to an increase in the average oxidation state of surface iron induced by $H_2O_2$, which promotes the formation of more oxidized Fe–O species and strengthens the Fe–O bonding environment, thus requiring higher binding energy for photoemission.

Regarding the magnesium XPS spectra (Fig. 10 (b)), the as-prepared nanoparticles display a single Mg 1$s$ peak centered at ~1304 eV, consistent with $Mg^{2+}$ in the spinel lattice. The NPs+MB sample shows no meaningful variation in either peak position or intensity, indicating that interaction with the dye does not alter the oxidation state or surface environment of Mg. In contrast, a clear decrease in Mg 1$s$ intensity is observed after treatment with $H_2O_2$. This attenuation is most likely associated with a reduction in the surface contribution of magnesium, which may arise from partial surface leaching, preferential surface enrichment of iron due to peroxide-driven redox processes, or the formation of surface hydroxylated layers that screen Mg from XPS detection. Importantly, the binding-energy position remains unchanged, confirming that Mg retains its divalent state despite these surface modifications [26]

In the O 1$s$ region, the main peak at 529.5–530.0 eV (Fig. 10 (c)) is assigned to lattice oxygen ($O^{2-}$) bonded to metal ions within the spinel structure. A secondary feature at 531–532 eV corresponds to non-structural or surface oxygen species, such as hydroxyl groups, adsorbed carbonates, or oxygen-containing intermediates formed during catalysis [26, 27]. In sample NPs+MB+$H_2O_2$, the O 1$s$ peak is broader and slightly shifted toward higher binding energies, indicating an increased presence of surface oxygen. This is consistent with oxidation induced by exposure to $H_2O_2$.

*3.5 Hydrodynamic evolution of MgFe$_2$O$_4$ nanoparticles under sonication*

Laser diffraction (LD) analyzer provides the hydrodynamic diameter of dispersed particles, allowing evaluation of aggregation state, colloidal stability, and dispersion quality during sonication. LD measurements show how the hydrodynamic behaviour of $MgFe_2O_4$ nanoparticles depends on ultrasonic power and exposure time (Fig. 11). At 3 W, the size distribution shifts progressively toward smaller diameters as sonication time increases. Large aggregates (80–100 μm) still present after 3 min are markedly reduced after 6–9 min, indicating that even mild ultrasound can fragment weakly bound clusters and improve dispersion.

Increasing the power to 7 W for 1 min produces a narrower distribution centered at ~40–60 μm, consistent with more intense cavitation capable of breaking compact aggregates and exposing additional reactive surfaces.

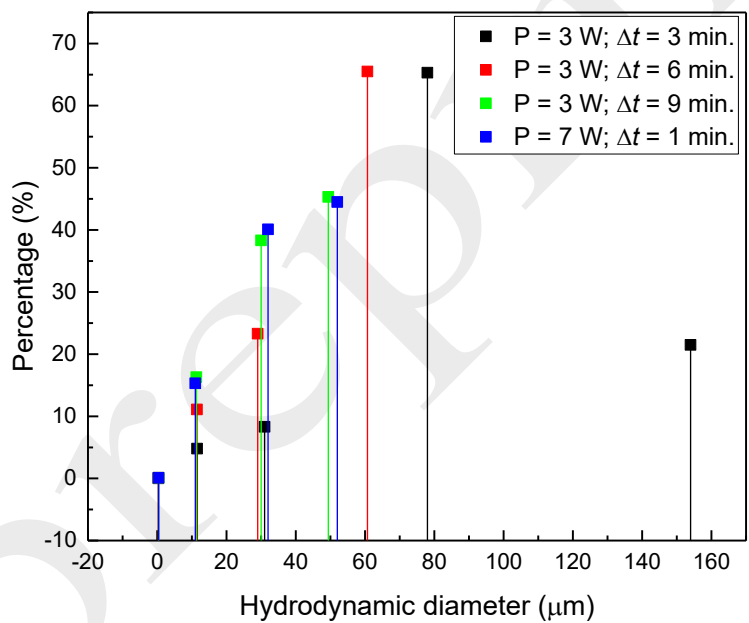

**Figure 11**: Hydrodynamic diameter size distribution of MgFe$_2$O$_4$ nanoparticles under different sonication power (P) and interval of time (Δt).

These hydrodynamic changes directly impact catalytic performance. Initially, large aggregates act as efficient cavitation nucleation centres, while continued sonication breaks them into smaller, more stable particles with greater accessibility of Fe$^{3+}$/Fe$^{2+}$ sites and improved mass transfer. Thus, sonication not only reduces the hydrodynamic diameter but also continuously reactivates the suspension through cavitation and shear.

Overall, suspensions with intermediate hydrodynamic sizes (tens of micrometres) offer the best balance between cavitation efficiency and surface availability. The LD results therefore highlight sonication as both a dispersive and a catalyst-activation process that dynamically optimizes the exposure of reactive sites during pollutant degradation.

## 4. Summary and Conclusions

The results of this study highlight the effectiveness and robustness of magnesium ferrite ($MgFe_2O_4$) nanoparticles synthesized via a simple autocombustion method as efficient catalysts for the degradation of methylene blue (MB) under mild and straightforward conditions. Structural and spectroscopic analyses confirm the formation of a stable spinel phase, predominantly featuring iron in the $Fe^{3+}$ state, with the crystalline framework maintaining its integrity after numerous catalytic cycles. This structural stability is vital for applications in environmental remediation, ensuring the durability and reusability of the catalyst without necessitating regeneration or complex processing steps.

Testing of the $MgFe_2O_4/H_2O_2$ system reveals that it achieves complete dye removal within approximately 30 minutes, even without illumination. This indicates that the catalytic process is primarily driven by surface-mediated Fenton-like reactions involving cycles of $Fe^{3+}/Fe^{2+}$ redox, with minimal enhancement under visible light, signifying an energy-efficient approach that operates effectively without artificial light or additional heating. The importance of nanoparticle dispersion is emphasized through experiments under various mixing conditions; ultrasonic treatment significantly enhances catalytic activity by facilitating contact between nanoparticles and reactants. The process of cavitation not only disperses the particles but also continuously reactivates their surfaces, leading to increased production of reactive oxygen species.

Further insights from FTIR and XPS analyses reveal that the characteristic tetrahedral and octahedral Fe–O vibrations persist during the catalytic process, confirming the integrity of the spinel lattice. Additionally, the emergence of oxygenated and sulfonate-related bands points to the temporary formation of oxidation intermediates, while the XPS data indicate an increase in surface oxygen species after exposure to $H_2O_2$, illustrating the efficient activation of hydrogen peroxide on the nanoparticle surface. This research demonstrates that $MgFe_2O_4$ possesses a chemically dynamic and regenerable surface, effectively activating peroxide without the need for dopants or composite supports.

Reusability tests show a marked difference in performance based on the presence of peroxide. Without $H_2O_2$, the degradation efficiency (DE) rapidly declines from 56% in the initial cycle to approximately 20% in subsequent cycles due to deactivation linked to surface fouling by dye molecules and oxidized intermediates. Conversely, the incorporation of $H_2O_2$ sustains high catalytic performance during the initial cycles, as peroxide facilitates continuous redox cycling and the oxidative removal of adsorbed organic species, thereby prolonging the availability of active sites and delaying surface passivation. However, a decrease in performance becomes evident in the third cycle, particularly in darkness, suggesting the potential onset of deeper surface modifications or partial aggregation of nanoparticles.

Moreover, the intrinsic magnetic separability of $MgFe_2O_4$ allows for quick recovery of the catalyst without complex filtration or centrifugation processes. When coupled with the low synthesis cost, structural resilience, and high catalytic efficiency, these attributes underscore the potential of the $MgFe_2O_4/H_2O_2$ system for decentralized and low-maintenance wastewater treatment applications, where simplicity, sustainability, and affordability are critical.

In summary, this study illustrates that $MgFe_2O_4$ nanoparticles are promising candidates for low-cost, low-maintenance water treatment solutions through their efficiency, stability, and magnetic recoverability, presenting a practical, scalable, and environmentally friendly approach to water purification technologies. The mechanistic and kinetic insights provided here emphasize the relevance of surface-mediated Fenton-like processes over purely photocatalytic pathways and contribute to the rational design of ferrite-based catalysts for sustainable water purification applications.

**Acknowledgements**

This work was supported with grants from CONICET (PIP 11220210100751CO), Agencia I+D+I (PICT-2020-SERIEA-00865), and UNLP (11/X953).

**Compliance with Ethical Standards:**

Conflict of Interest: The authors declare that they have no conflict of interest.

**Declaration of generative AI and AI-assisted technologies in the writing process**

During the preparation of this work the authors used OPENAI/CHATGPT to improve language and readability. After using this tool/service, the authors reviewed and edited the content as needed and take full responsibility for the content of the publication.

# Supplementary Information

The following material complements the results presented in the article **Magnetically recoverable MgFe$_2$O$_4$ nanoparticles as efficient catalysts for rapid dye degradation in water,** by A.F. Cabrera, C. E Rodríguez Torres, P. M. de la Presa and S. J. Stewart

## X-ray diffraction results

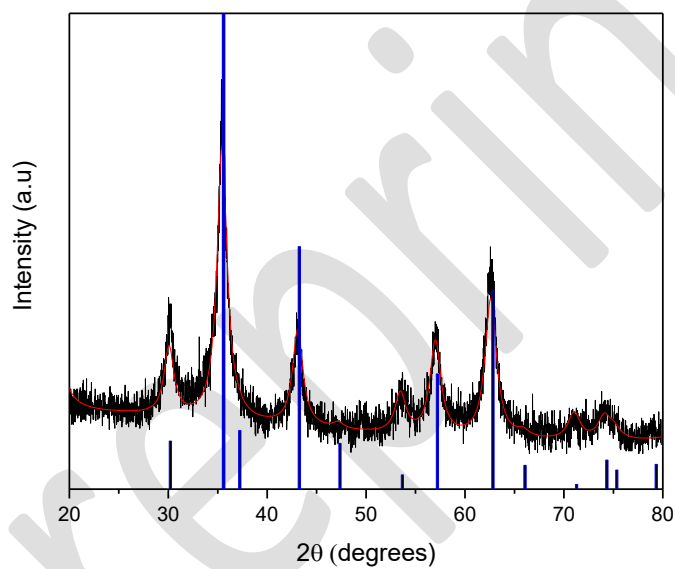

**Figure SI-1**: X-ray diffraction pattern (black) and Rietveld refinement of the powder data performed using the *MAUD* software (red line). Reflections of the spinel structure are indicated.

The sample was characterized by X-ray diffraction (XRD). The analysis of the pattern (red line) was performed using the *MAUD* software [1], assuming the crystallographic structure of magnesium ferrite with space group Fd3m (Crystallography Open Database file COD 1011241.cif). The refinement yielded a lattice parameter of 8.391(4) Å and a crystallite size of approximately 89(1) Å, in good agreement with the value obtained from transmission electron microscopy (TEM) measurement (see below).

**TEM and HRTEM results**

The as-prepared MgFe$_2$O$_4$ NPs and NPs after being used for discoloration with peroxide were analyzed using High Resolution Transmission Electron Microscopy (HRTEM) with a JEOL JEM 2100TM to determine the morphology and size distribution of the NPs. The size distribution of NPs was analyzed using ImageJ software.

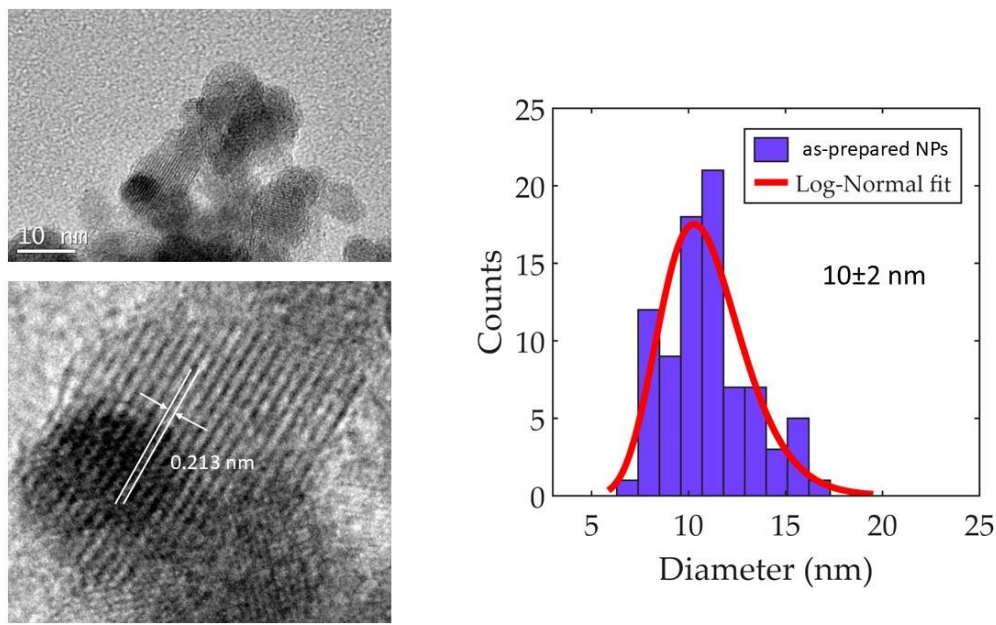

**Figure SI-2**: TEM and HRTEM micrography of as-prepared MgFe$_2$O$_4$ before its use as catalyst (left). The distribution od particle sizes is also shown (right).

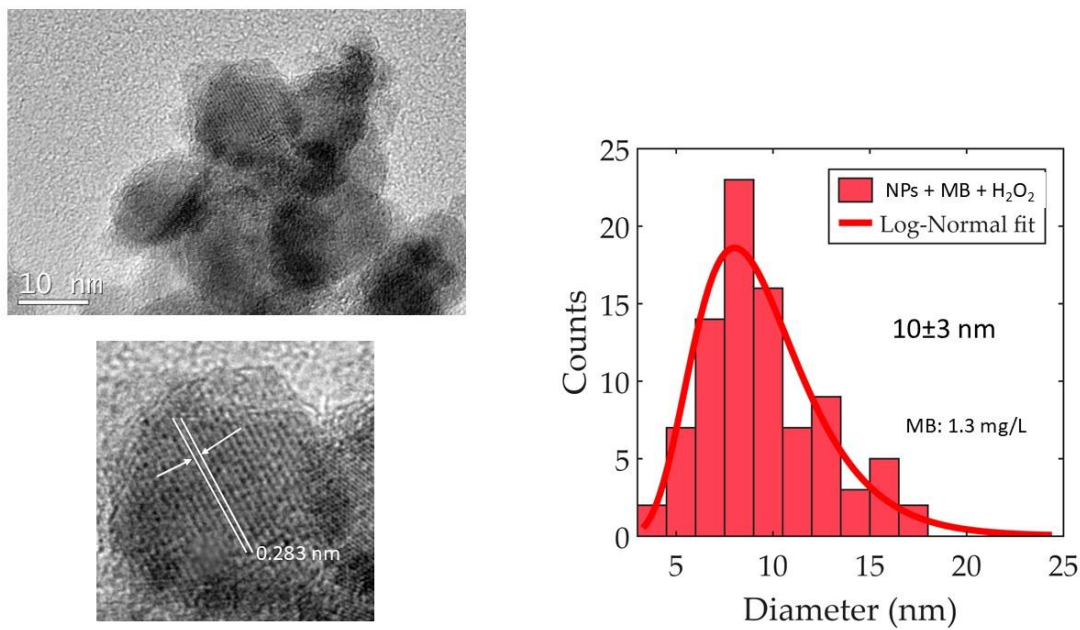

**Figure SI-3**: TEM and HRTEM micrography of as-prepared MgFe$_2$O$_4$ after used as catalyst (left) in the degradation of methylene blue (test F, system MB + NPs + H$_2$O$_2$, with 1.3 mg.L$^{-1}$ of MB). The distribution of particle sizes is also shown (right).

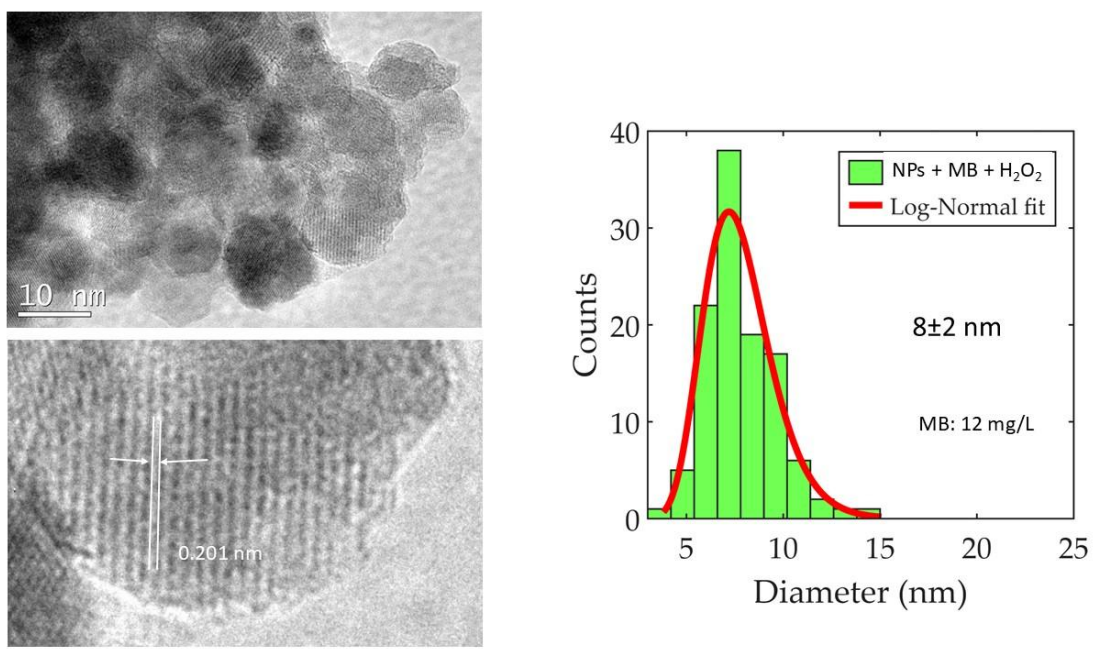

**Figure SI-4**: TEM and HRTEM micrography of as-prepared MgFe$_2$O$_4$ after used as catalyst (left) in the degradation of methylene blue (test F, system MB + NPs + H$_2$O$_2$, with 12 mg.L$^{-1}$ of MB). The distribution of particle sizes is also shown (right).

## Mössbauer spectroscopy results

Mössbauer spectroscopy was carried out at room temperature to examine the oxidation state and local environment of Fe ions in MgFe$_2$O$_4$. The spectrum (Figure SI-5) is dominated by doublet components attributed to Fe$^{3+}$ species in paramagnetic or superparamagnetic environments, with no evidence of Fe$^{2+}$ ions. A minor sextet contribution is also detected.

Spectral fitting using the *Recoil* program [2] (parameters summarized in Table I) assigns the sextet to Fe$^{3+}$ ions in the spinel structure. The doublets exhibit an isomer shift of ~0.35 mm s$^{-1}$ and different quadrupole splittings, consistent with Fe$^{3+}$ ions in distorted tetrahedral and octahedral sites. These results confirm the inverse spinel configuration of MgFe$_2$O$_4$ and the presence of multiple Fe$^{3+}$ environments.

| site | H (T) or ΔQ(mm/s) | σ(mm/s) | δ (mm/s) | Relative Area |
|---|---|---|---|---|
| Doublet 1 | 0.66$_1$ | 0.42$_1$ | 0.327$_2$ | 79$_1$ |
| Doublet 2 | 1.98$_7$ | 0.8 | 0.36$_2$ | 19$_2$ |
| Sextet | 46.5$_3$ | | 0.44$_4$ | 2.0$_6$ |

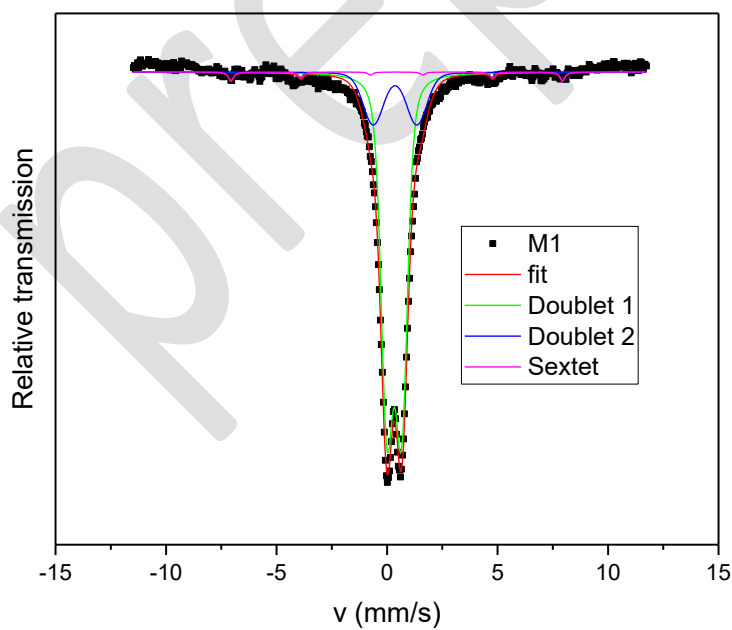

**Figure SI-5:** Mössbauer spectrum of MgFe$_2$O$_4$ as- prepared sample (NPs).

## Magnetic characterization

The magnetic behavior of the MgFe$_2$O$_4$ sample was examined by measuring the magnetization ($M$) as a function of the applied magnetic field ($H$) at room temperature. The M–H curve, shown in Figure SI-6, displays a typical *S-shaped* hysteresis loop with a narrow coercive field, indicating soft ferrimagnetic behavior. The magnetization increases rapidly at low fields and gradually approaches saturation at higher fields, reaching a maximum value of approximately 11 emu/g at 2 T. The small but finite coercivity and remanent magnetization reveal the presence of ferrimagnetic ordering, characteristic of spinel ferrites where Fe$^{3+}$ ions occupy both tetrahedral (A) and octahedral (B) sites. The relatively low coercive field suggests that the material possesses low magnetic anisotropy and that domain wall motion occurs easily under an external field. Overall, the magnetic response confirms that MgFe$_2$O$_4$ exhibits soft ferrimagnetic characteristics at room temperature, consistent with the cation distribution typical of spinel ferrites.

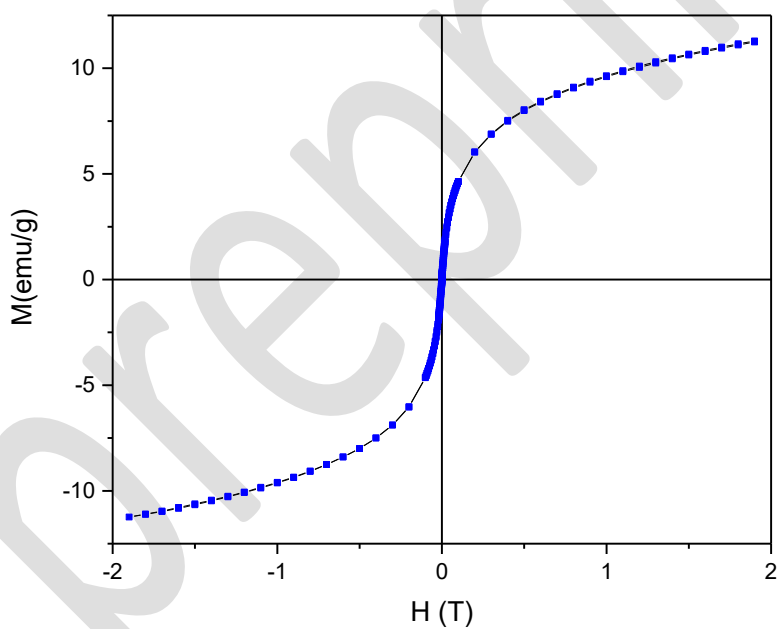

**Figure SI-6:** Magnetization (M) as a function of applied magnetic field (H) for the as-prepared MgFe$_2$O$_4$ sample, measured at room temperature.

## UV-visible spectroscopy results

To determine the optical band gap, the absorption coefficient (α) was calculated from the measured absorbance, and a Tauc plot of $(\alpha h\nu)^n$ versus photon energy $(h\nu)$ was constructed [3,4]. The relationship between the absorption coefficient and photon energy is given by the Tauc equation:

$$(\alpha h\nu)^n = C(h\nu - E_g)$$

where $C$ is a proportionality constant, $E_g$ is the optical band-gap energy, and $n$ depends on the nature of the electronic transition. Assuming a direct allowed transition ($n = 2$), the linear portion of the Tauc plot was extrapolated to the energy axis, yielding a band-gap value of 2.11 eV.

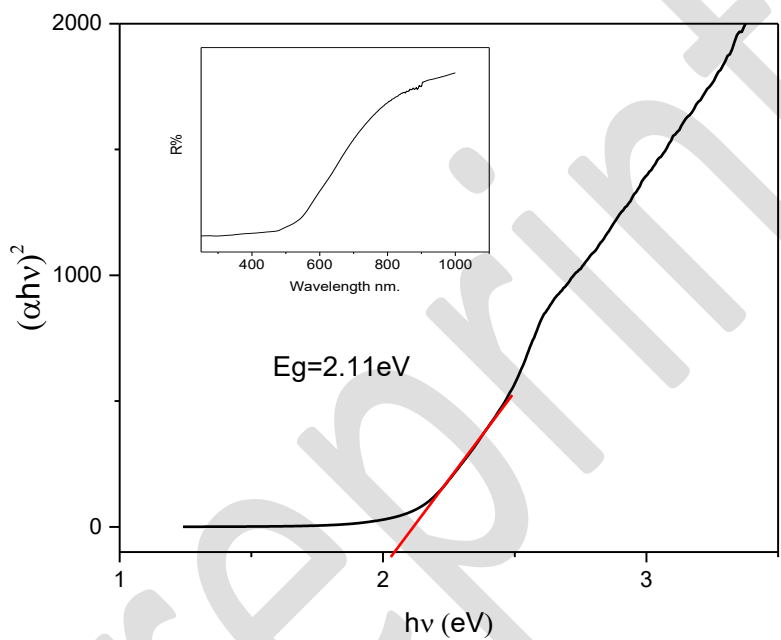

**Figure SI-7**: Tauc plot for estimating the optical band-gap energy [3,4] from absorbance data. Inset: diffuse reflectance spectrum versus wavelength

## References

[1] L. Lutterotti, P. Scardi, Simultaneous structure and size–strain refinement by the Rietveld method, J. Appl. Crystallogr. 23 (1990) 246–252, https://doi.org/10.1107/S0021889890002382.

[2] D.G. Rancourt, J.Y. Ping, Voigt-based methods for arbitrary-shape static hyperfine parameter distributions in Mossbauer ̈spectroscopy, Nucl. Instrum. Methods Phys.Res. B 58 (1991) 85–97, https://doi.org/10.1016/0168-583X(91)95681-3.

[3] P. Kubelka, Ein Beitrag zur Optik der Farbanstriche (Contribution to the optic of paint), Zeitschrift fur technische Physik 12 (1931) 593–601.

[4] M. Nowak, B. Kauch, P. Szperlich, Determination of energy band gap of nano-crystalline SbSI using diffuse reflectance spectroscopy, Rev. Sci. Instrum. 80 (4) (2009) 046107, https://doi.org/10.1063/1.3103603.